# Origin of subgap states in normal-insulator-superconductor van der Waals heterostructures

Paritosh Karnatak[1*†], Zarina Mingazheva[1*], Kenji Watanabe[2], Takashi Taniguchi[2], Helmuth Berger[3], László Forró[3,4], Christian Schönenberger[1,5]
[1] *Department of Physics, University of Basel, Basel, Switzerland*
[2] *Research Center for Functional Materials, National Institute for Material Science, 1-1 Namiki, Tsukuba 305-0044, Japan*
[3] *Institute of Condensed Matter Physics, Ecole Polytechnique Fédérale de Lausanne, Lausanne, Switzerland*
[4] *Stavropoulos Center for Complex Quantum Matter, Department of Physics, University of Notre Dame, Notre Dame, IN, USA and*
[5] *Swiss Nanoscience Institute, University of Basel, Basel, Switzerland*

Superconductivity in van der Waals materials, such as $NbSe_2$ and $TaS_2$, is fundamentally novel due to the effects of dimensionality, crystal symmetries, and strong spin-orbit coupling. In this work we perform tunnel spectroscopy on $NbSe_2$ by utilizing $MoS_2$ or hexagonal Boron Nitride (hBN) as a tunnel barrier. We observe subgap excitations and probe their origin by studying various heterostructure designs. We show that the edge of $NbSe_2$ hosts many defect states, which strongly couple to the superconductor and form Andreev bound states. Furthermore, by isolating the $NbSe_2$ edge we show that the subgap states are ubiquitous in $MoS_2$ tunnel barriers, but absent in hBN tunnel barriers, suggesting defects in $MoS_2$ as their origin. Their magnetic nature reveals a singlet or a doublet type ground state and based on nearly vanishing g-factors or avoided-crossing of subgap excitations we highlight the role of strong spin-orbit coupling.

Superconductivity in the two-dimensional limit is driven by a unique interplay between dimensionality, crystal symmetries, correlated electron effects and, if present, the role of spin-orbit coupling. This often results in various competing ground states and gives rise to rich novel phenomena. Ultimately two-dimensional van der Waals superconductors are illustrative examples. Naturally superconducting $NbSe_2$ and $TaS_2$ have been recently isolated and studied [1, 2], and $MoS_2$ has been doped into a superconducting state [3]. In their monolayer or few-layer forms these van der Waals superconductors display novel phenomena, such as the survival of superconductivity up to tens of Teslas of applied in-plane magnetic field [1, 2], layer dependent superconducting properties [2] and competition with other phases [4]. Furthermore, it is predicted that these materials can be externally tuned to host novel topological phases [5, 6] and there are expectations of the presence of unconventional pairing mechanisms in Ising superconductors [7].

These features essentially result from the large spin-orbit coupling (SOC) and the crystal symmetry in these materials. For this SOC, called the Ising-type, the corresponding spin orbit magnetic field points out-of-plane and in opposite directions in the opposite valleys of the hexagonal Brillouin zone of these materials [1, 3]. This splits the spin degenerate bands and the majority singlet Cooper pairs are expected to be formed from opposite valleys. As the large spin orbit magnetic field (some estimates indicate $B_{so} \sim 100$ T [3]) pins the spins out-of-plane, an applied in-plane magnetic field (usually smaller than $B_{so}$) hardly affects the electron spins and thus the Cooper pairs survive large Zeeman fields.

Recently, proximity induced superconductivity in semiconducting nanostructures has been widely investigated [8–12], primarily driven by the proposals for topological quantum computation [13, 14]. Additionally, low dimensional structures coupled to van der Waals superconductors with large SOC provide a rich platform to investigate the nature of Andreev bound states. It may also offer insights into the unconventional superconducting properties. In this regard, tunnel spectroscopy is a versatile tool to probe the superconducting density of states (DOS). Electronically gapped van der Waals materials provide high quality tunnel barriers that allow an unprecedented control over the barrier thickness and the interface quality. They are also especially well suited to probe the air-sensitive van der Waals superconductors. Tunnel spectroscopy in such heterostructures has revealed the presence of Andreev levels in the subgap spectrum [15, 16]. However, the exact origin and nature of these bound states has not been systematically investigated, and it is not known if such bound states reside in the tunnel barriers or are hosted on the $NbSe_2$ surface [17–19]. The role of spin-orbit coupling in determining the Andreev level ground state and their magnetic nature also remains to be understood.

In this work we perform tunneling spectroscopy on $NbSe_2$ by utilizing $MoS_2$ or hexagonal Boron Nitride (hBN) [20] as a tunnel barrier and Ti/Au as the normal leads. We find that the single particle gapped spectrum is often interrupted by the presence of subgap excitations and we probe their origin by studying various heterostructure designs. We show that the edge of $NbSe_2$ hosts many defect states, some of which are strongly coupled to the superconductor. However, we also observe subgap excitations in devices where the $NbSe_2$ edge is electrically isolated. We show that these subgap excitations arise from defects in $MoS_2$ and are absent in hBN tunnel barriers. We probe the magnetic nature of these subgap states by studying their evolution in applied magnetic fields and reveal the nature of ground states, as well as highlight the role of spin-orbit coupling.

*Equal Contributions
†Correspondence to: paritosh.karnatak@unibas.ch

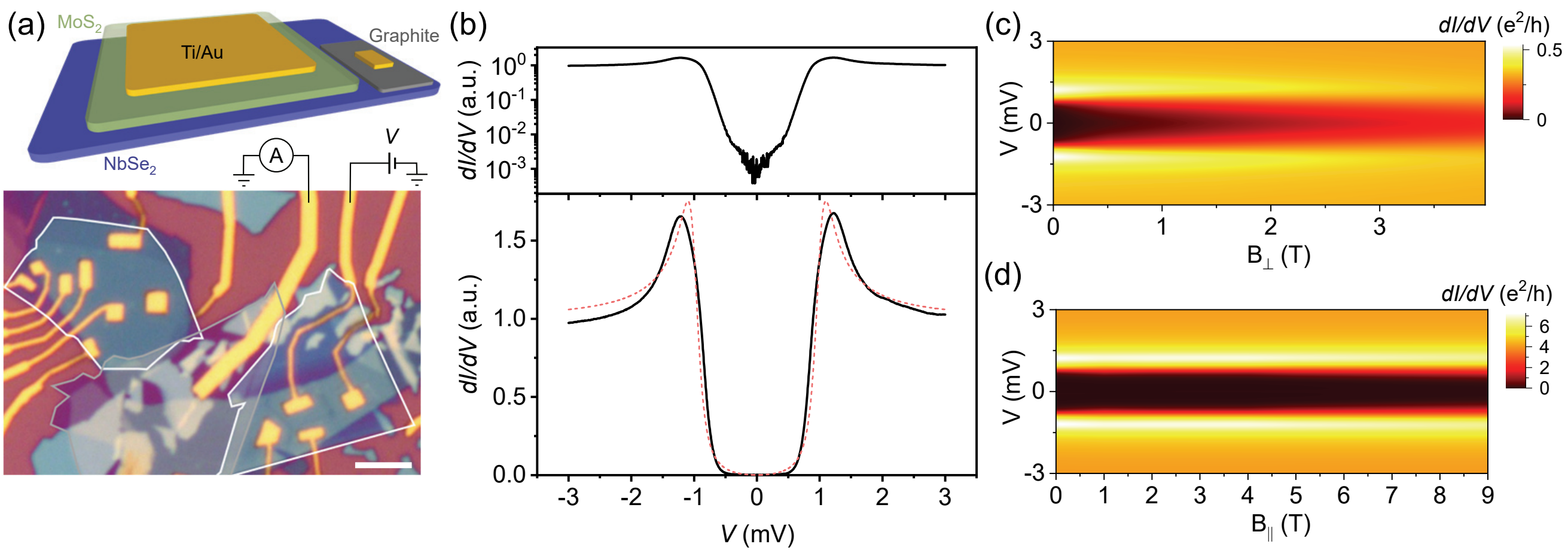

Figure 1: Device design for planar tunneling devices and differential conductance measurements. (a) Schematic shows the normal-insulator-superconductor junctions formed by depositing Ti/Au on the $MoS_2$ (or hBN) and $NbSe_2$ stack. The optical image shows a typical device with $MoS_2$ (white outline) and graphite (grey outline) transferred on $NbSe_2$ crystal (blueish color). Scale bar is 5 $\mu$m. (b) $dI/dV$ measured shows a hard superconducting gap with a suppression factor $G_N/G_0 \sim 800$. The dashed red curve is eq 1 with the parameters $\Delta \sim 1.0$ meV and $\Gamma \sim 0.11$ meV. (c) The out-of-plane magnetic field leads to the softening of the superconducting gap. (d) $dI/dV$ measured in an in-plane magnetic field shows that the superconducting gap is robust up to 9 T.

The normal-insulator-superconductor (NIS) type planar tunnel junctions are fabricated by stacking $MoS_2$ (3−5 layers) or hBN (3 layers) on $NbSe_2$ crystals ($\sim$ 3 nm −20 nm) in a glovebox in $N_2$ atmosphere. $MoS_2$ or hBN act as the tunnel barrier and prevent $NbSe_2$ from oxidation, see schematic and a representative device image in Figure 1a. We have studied 8 devices and over 50 tunnel junctions and a summary of results is presented here. Further details of fabrication, device parameters and measurements can be found in the Supporting Information (SI).

The differential conductance across an NIS junction can be written as [21, 22]

$$\frac{dI}{dV} \propto \int_{-\infty}^{+\infty} N_S(E,\Gamma,\Delta)\frac{df(E-eV,T)}{dV}dE \qquad (1)$$

where $N_S$ is the single particle DOS as a function of energy $E$ for the superconducting electrode with a superconducting gap $\Delta$ and a broadening parameter $\Gamma$; $f(E,T)$ is the Fermi- Dirac distribution at a finite temperature $T$ and $V$ is the bias voltage applied across the tunnel barrier. The superconducting DOS can be modeled by the Dynes formula [23]

$$N_S(E,\Gamma,\Delta) = \mathrm{Re}\left\{\frac{E-i\Gamma}{\sqrt{(E-i\Gamma)^2-\Delta^2}}\right\} \qquad (2)$$

It is instructive to see that at zero temperature eq 1 reduces to $dI/dV \propto N_S(eV,\Gamma,\Delta)$ and the differential conductance measurement across a tunnel barrier probes the DOS of the superconductor. At finite temperature the DOS features are broadened by $\sim k_BT$. One such measurement is shown in Figure 1b, with a well-defined superconducting gap and a suppression factor $G_N/G_0 \gtrsim 800$, where $G_0$ and $G_N$ are the differential conductance in the superconducting gap (at $V = 0$) and outside the gap (typically $V \sim 3$ mV) respectively, emphasised in log-scale in the top panel of Figure 1b. We typically observe hard gaps across our tunnel barriers with $G_N/G_0 \gtrsim 100$, indicating high quality tunnel barriers and consequently the suppression of Andreev processes. A plot of eq 1 with a gap of $\Delta \sim 1.0$ meV and broadening parameter $\Gamma \sim 0.11$ meV is shown in Figure 1b. Moreover, $dI/dV$ measurements performed in a perpendicular magnetic field reveal significant softening of the superconducting gap, see Figure 1c, as expected for $NbSe_2$ at this scale due to the orbital depairing [24, 25]. However, the Ising protection against an applied in-plane magnetic field is noticeable for $NbSe_2$, as seen in Figure 1d. We observe that the superconducting gap is robust ($G_N/G_0 \sim 100$) up to 9 T, only limited by the cryostat magnet. This allows us to study the behavior of the subgap states in a large (in-plane) magnetic field, as discussed later.

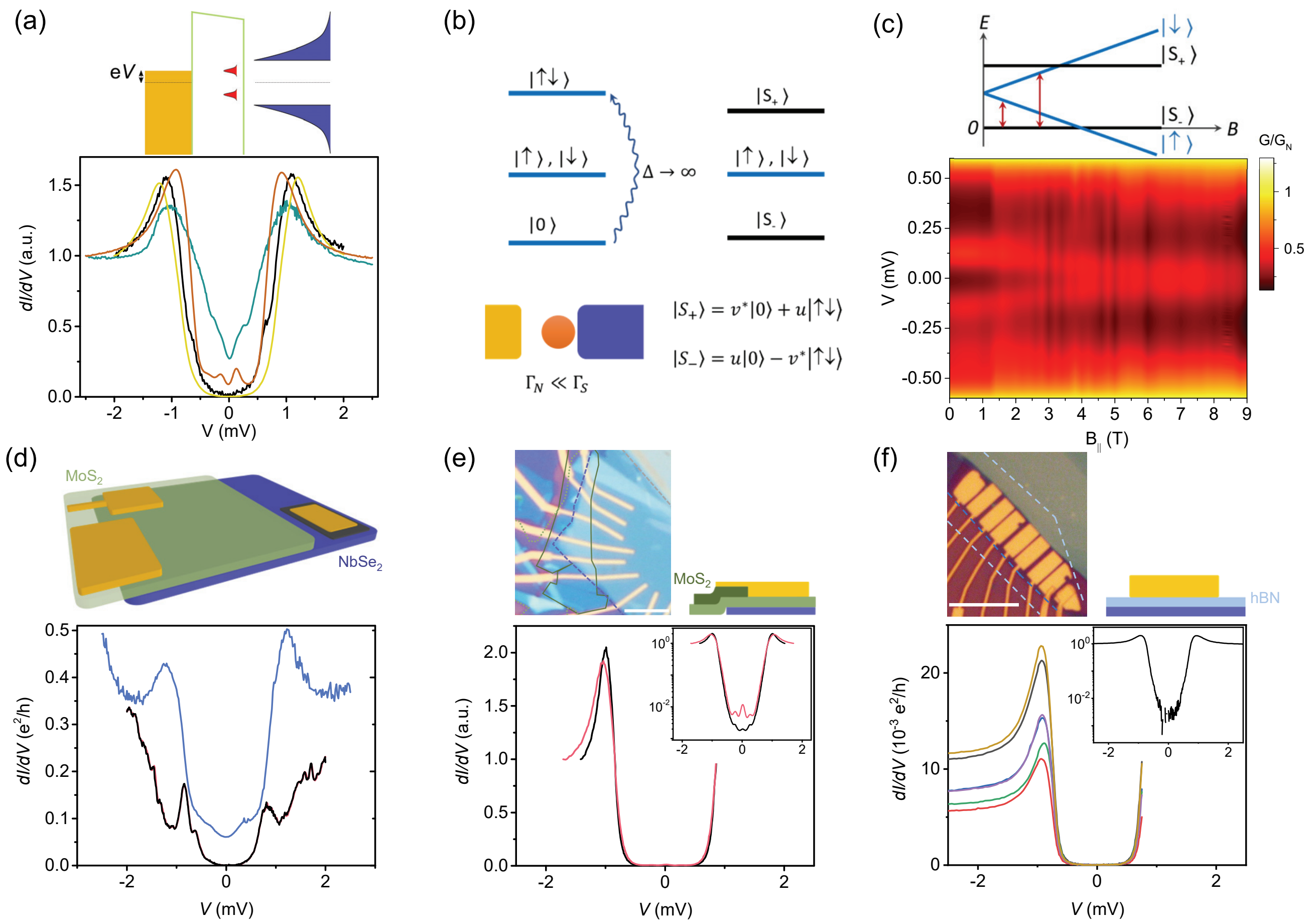

Figure 2: Origin of subgap states. (a) $dI/dV$ measurements show the presence of subgap excitations. Schematic shows a possible mechanism where a defect strongly couples to the superconductor. (b) The electronic states in a quantum dot are modified when it is strongly coupled to the superconductor. (c) Magnetic field evolution of the subgap excitations indicates that at $B = 0$ the ground state is a singlet but the system undergoes a quantum phase transition to a doublet ground state at a finite magnetic field. The jumps in magnetic field are a result of imperfect alignment of the $B_{\|}$ that leads to discrete units of flux entering the tunnel junction area. (d) $dI/dV$ measurements on the edge of $NbSe_2$. Repeated runs are shown by black and red lines (barely visible, lie on top of each other), while another edge contact is shown in blue. (e) $dI/dV$ measurements with the edge of $NbSe_2$ electrically isolated by using a thicker $MoS_2$ (solid green outline) at the edge of $NbSe_2$ (blue dashed line). Scale bar is 5 $\mu$m. Inset in a log-scale highlights the presence of subgap excitations. (f) $dI/dV$ measurements for hBN used as a tunnel barrier, shows absence of subgap excitations. Inset in log-scale shows the absence of subgap excitation down to the measurement noise floor. Scale bar in the optical image is 10 $\mu$m.

Unlike the spectrum shown in Figure 1b, however, we often observe discrete subgap features in $MoS_2$ tunnel barrier junctions, see Figure 2a. Such subgap features can result from discrete electronic states in the tunnel path, modified by the superconducting proximity effect. The discrete states themselves may arise from a defect or an impurity in the tunnel barrier or on the surface of the superconductor [17–19]. The formation of such Andreev levels has recently been widely explored, especially in semiconducting nanowires coupled to superconductors, and is fairly well understood [9, 11, 12]. We model the defect state as a quantum dot coupled to a superconductor. A spin degenerate, single orbital level in an isolated quantum dot has four eigenstates $|0\rangle$, $|\uparrow\rangle$, $|\downarrow\rangle$ and $|\uparrow\downarrow\rangle$. When the quantum dot couples to a superconductor, the empty $|0\rangle$ and the doubly occupied states $|\uparrow\downarrow\rangle$ are hybridised via virtual Andreev processes that exchange two electrons with the dot. If the quasiparticles in the superconductor can be neglected ($\Delta \to \infty$, the so called superconducting atomic limit), this hybridisation results in two BCS-like singlet eigenstates $|S_-\rangle = u\,|0\rangle - v^*\,|\uparrow\downarrow\rangle$ and $|S_+\rangle = v\,|0\rangle + u^*\,|\uparrow\downarrow\rangle$, given by the Bogoliubov-de Gennes (BdG) transformation [26–28], where $u$ and $v$ are the BdG amplitudes. For single average occupancy of the dot, the system has two possible ground states - either the degenerate doublet $|\uparrow\rangle$, $|\downarrow\rangle$ or the singlet eigenstate

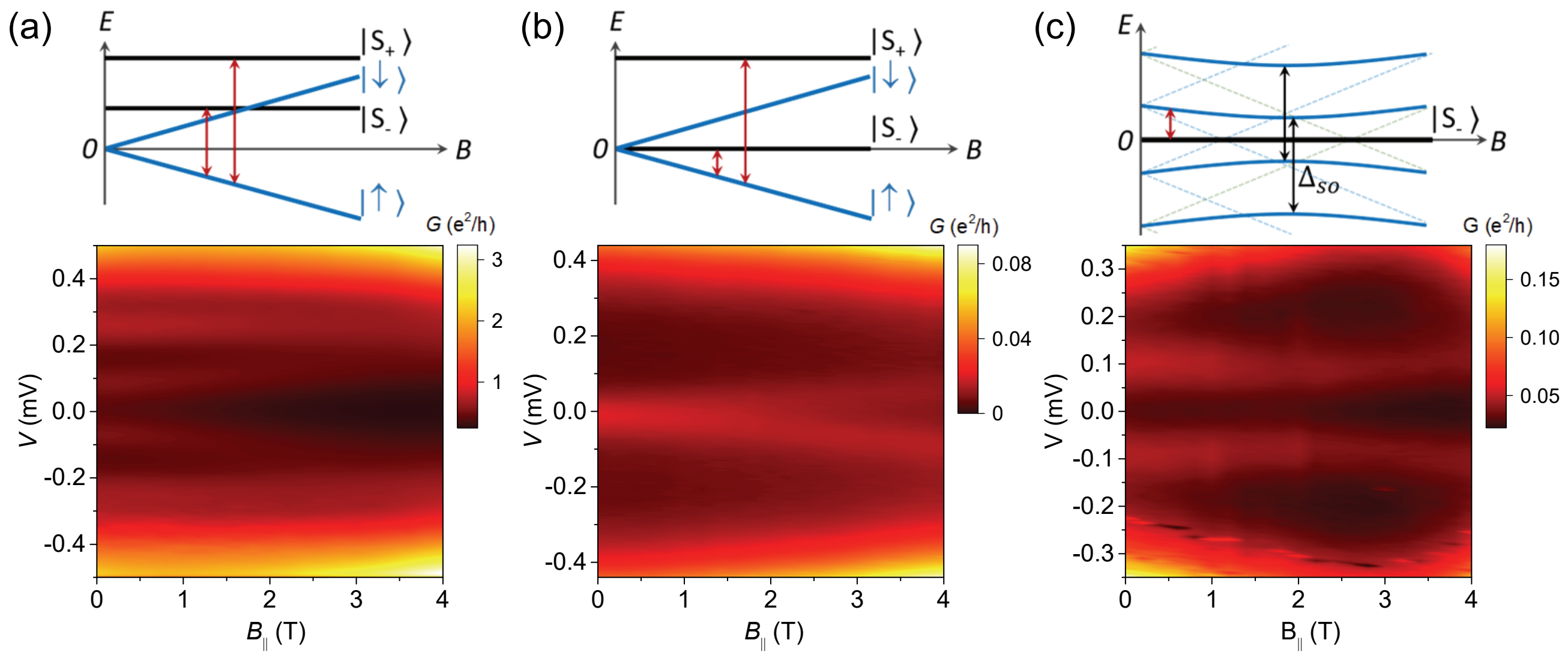

Figure 3: Anomalous subgap excitations. (a) Multiple subgap states are seen at $B = 0$, they evolve with an applied in-plane magnetic field but with a different g-factor. (b) $dI/dV$ measurements show a zero bias excitation at $B = 0$, which splits in magnetic field. The transition to the higher singlet is likely at the gap edge and is not visible in our measurement. (c) Subgap excitations show an avoided crossing feature with a minimal splitting of $\approx 0.185$ meV at $B_{||} \approx 2$ T.

$|S_-\rangle$. The energy of the singlet states is given by [27]

$$E_\pm = U/2 \pm \sqrt{\xi_d^2 + \Gamma_s^2} + \xi_d$$

where $U$ is the charging energy, $\xi_d = \epsilon_d + U/2$ with $\epsilon_d$ being the bare energy of the doublet states $\{|\uparrow\rangle, |\downarrow\rangle\}$ and $\Gamma_s$ the coupling to the superconductor. Therefore, the competition between the ground states $\{|\uparrow\rangle, |\downarrow\rangle\}$ and $|S_-\rangle$ depends on the relative magnitudes of various energy scales in the system. In general, a stronger coupling $\Gamma_s$ to the superconductor favors a singlet $|S_-\rangle$ ground state whereas a large charging energy $U$ results in a doublet ground state $\{|\uparrow\rangle, |\downarrow\rangle\}$. In this work, we discuss the subgap features in terms of the Andreev bound states and this framework holds when the quasiparticles in the superconductor do not play a role. However, in principle our experiments cannot distinguish if the singlet is the superposition of $|0\rangle$ and $|\uparrow\downarrow\rangle$ (Andreev bound state) or is formed between one electron on the dot and another on the superconductor (Yu-Shiba-Rushinov state) [29–31]. Quasiparticles in the superconductor could play a role if $\Delta \sim \Gamma_s$.

Applying a dc voltage bias $V$ across the tunnel barrier, that is equivalent to the energy of the excited state, results in the transfer of an electron into the dot from the normal lead. This electron can form a Cooper pair to enter the superconductor and consequently a hole is retro-reflected into the normal lead. Symmetric across the Fermi energy a time-reversed process occurs and can be observed as a similar feature at the opposite dc voltage bias. Thus, the electron-hole symmetric subgap features in tunnel spectroscopy probe the excitation energy of the subgap states. An external magnetic field causes Zeeman splitting of the spin degenerate doublet states and provides a key tool to study the nature of the localised ground states, see Figure 2c and the discussion later.

We first investigate the origin of such subgap excitations - whether they reside in the tunnel barrier or on the surface of the superconductor [17–19]. We notice that the tunnel junctions with a large overlap with the edge of the $NbSe_2$ crystal exhibit multiple features in $dI/dV$ both outside and inside the gap, see Figure 2d. The repeatability of these features in multiple sweeps indicates that they represent discrete energy levels and do not arise from time dependent noise. They likely arise from defects present at the $NbSe_2$ edge cleaved during exfoliation, some of which strongly couple to the superconductor and show up as subgap excitations. While it may not be surprising that the $NbSe_2$ edge hosts many defects, this may be critical for the topological edge states predicted in $NbSe_2$ with an applied in-plane magnetic field [5, 6]. Instead, it would be crucial to engineer the boundary of the topological and the trivial phase on the bulk of $NbSe_2$, as in a recent study [32].

In a simple planar tunnel junction a part of the normal 'wire' always crosses the $NbSe_2$ edge, see Figure 2d schematic. Therefore, next we address the question - if all the subgap states that we observe arise from such defect states at the edge of $NbSe_2$. We do this by electrically isolating the $NbSe_2$ edge by transferring additional $MoS_2$ layers over the edge of $NbSe_2$, see the optical image and the schematic of Figure 2e. The corresponding $dI/dV$ curves plotted in Figure 2e exhibit a well-behaved superconducting gap. The subgap states are now rare, but still present in multiple junctions, as shown in the inset. This points to other source(s) of defect states, in addition to those at the edge of the $NbSe_2$ crystal. The possibilities that remain are the defect states in the tunnel barrier or on the surface of the superconductor.

To address this, we replace the $MoS_2$ tunnel barrier with 3 layers of hBN, known to be an effective tunnel barrier. In particular, the defect density in hBN is small [20, 33] and likely three orders of magnitude smaller than that in $MoS_2$ [34–37], although we are not aware of direct comparative studies. The differential conductance for six such tunnel junctions, each with an area $\sim 10\ \mu\text{m}^2$, is shown in Figure 2f. While tunnel spectroscopy shows a well behaved superconducting gap with a suppression factor $G_N/G_0 \sim 300$, we do not observe subgap features in any hBN tunnel junction down to our measurement resolution, as evident from the log-scale plot in the inset of Figure 2f. This leads us to believe that the subgap features in $MoS_2$ / $NbSe_2$ tunnel junctions arise either from the edge of the $NbSe_2$ crystal or defects in $MoS_2$ that strongly couple to the superconductor.

Further, we study the subgap excitation spectrum in an applied in-plane magnetic field. The Zeeman splitting of the doublet states $\{|\uparrow\rangle, |\downarrow\rangle\}$ results in unique features in the excitation spectrum which allows the identification of the ground state. One such measurement is shown in Figure 2c, where at $B_{\parallel} = 0$ two subgap excitations are visible at $V \approx \pm 0.13$ mV. With an applied in-plane magnetic field $B_{\parallel}$ the subgap features split (effective g-factor of $\sim 0.7$), where one branch moves towards zero bias and the other (weakly visible for $V > 0$) moves towards the gap edge (see SI for the second derivative). The overall behavior can be understood by considering that the dot is in a singlet $|S_-\rangle$ ground state at $B_{\parallel} = 0$. At $V \approx 0.13$ mV, the chemical potential of the normal lead is aligned to the spin degenerate doublet excited state. With increasing $B_{\parallel}$, the doublet splits resulting in the excitation energy to the lower branch decreasing while the excitation energy to the upper branch increases, as illustrated in the Figure 2c schematic. In fact, for $B_{\parallel} > 6$ T when the lower branch crosses zero energy, the system undergoes a quantum phase transition and the ground state changes to the doublet ground state. See SI for another such example. The appearance that the bound state sticks to zero energy for $B_{\parallel} > 6$ T, is either the result of two wide ($FWHM \sim 0.18$ meV) bound states crossing or the influence of spin-orbit mixing with higher orbital levels, as discussed later.

The ground state of the dot coupled to a superconductor depends on the relative strengths of various energy scales - the tunnel coupling of the dot to the superconductor $\Gamma_s$, the charging energy $U$, the superconducting gap $\Delta$, and the energy of the dot level relative to the chemical potential of the superconductor $\xi_d$ . Since a finite $\Gamma_s$ is necessary for the visibility of the subgap excitations and a large $\Gamma_s$ favors a singlet ground state, we observe singlet states nearly six times as frequently as doublet ground states (see SI for a count of ground states). One such case is shown in Figure 3a, where the excitations at $\approx \pm 0.08$ meV ($B_{\parallel} = 0$) move to higher absolute energies with an applied $B_{\parallel}$, expected for a doublet ground state. The schematic in Figure 3a demonstrates the mechanism. The subgap excitation visible at higher energies $\approx \pm 0.25$ meV ($B_{\parallel} = 0$) may be attributed to the transition to the higher singlet. But this is unlikely due to a different g-factor. Instead, this may result from another parallel Andreev bound state formed via a second defect, in a junction of size $\approx 3.5\ \mu\text{m}^2$, and large SOC may result in a nearly vanishing g-factor as discussed later. Rarely, a zero-bias peak is also observed at $B_{\parallel} = 0$ and we believe this results from an accidental degeneracy of the doublet and the lower singlet $|S_-\rangle$. One such spectrum along with the excitation energy schematic is shown in Figure 3b, where a zero-bias peak is observed for $B_{\parallel} = 0$ and splits for finite $B_{\parallel}$.

Finally, an avoided-crossing like feature, is shown in Figure 3c where the subgap excitations move towards zero bias but at $B_{\parallel} \approx 2$ T they start to move to higher absolute energies. We attribute this to the spin mixing and hybridisation of the doublet states that arise from higher orbital levels, due to SOC in the host material [38], as illustrated in Figure 3c schematic. The size of the splitting depends on the details of the defect which determine the strength of SOC and the relative directions of $B_{SO}$ and $B_{\parallel}$. No hybridisation occurs when the externally applied magnetic field is parallel to the internal spin orbit field [39, 40]. This may explain why splitting is not observed in other junctions. A large spin-orbit gap (compared to the doublet excitation energy), would also result in a reduced effective g-factor (see also Figure 3a).

In conclusion, we have performed tunnel spectroscopy on $NbSe_2$ by utilizing $MoS_2$ or hexagonal Boron Nitride (hBN) as a tunnel barrier and Ti/Au as the normal leads. We find that the single particle gapped spectrum often exhibits the presence of subgap excitations and we probe their origin by studying various heterostructure designs. We show that the edge of $NbSe_2$ hosts many defect states, some of which are strongly coupled to the superconductor. However, we also observe subgap excitations in devices where the $NbSe_2$ edge is electrically isolated. We show that while the subgap excitations are fairly ubiquitous in $MoS_2$ tunnel barriers they are absent in hBN tunnel barriers, suggesting that these subgap excitations arise from the defects in $MoS_2$. The evolution of subgap excitations in an applied in-plane magnetic fields allows us to probe the magnetic nature of the participating subgap states and reveals the nature of subgap ground states. Subgap excitations that anti-cross or show no dispersion with the Zeeman field highlight the role of spin-orbit coupling in the system.

P.K. and Z.M. contributed equally to this work. We thank Hadar Steinberg, Andreas Baumgartner and Je-lena Klinovaja for fruitful discussions. This project has received funding from the European Research Council (ERC) under the European Union's Horizon 2020 research and innovation programme: grant agreement No 787414 TopSupra, by the Swiss National Science Foundation through the National Centre of Competence in Research Quantum Science and Technology (QSIT), and by the Swiss Nanoscience Institute (SNI). K.W. and T.T. acknowledge support from the Elemental Strategy Initiative conducted by MEXT, Japan and the CREST (JPMJCR15F3), JST. P.K. and C.S. designed the experiments. P.K. and Z.M. fabricated the devices and performed the measurements. With inputs from C.S., P.K. and Z.M. performed the data analysis. H.B. and L.F. provided $NbSe_2$ crystals. K.W. and T.T. provided hBN crystals. P.K. and Z.M. wrote the manuscript with inputs from all authors. All data in this publication are available in numerical form at: https://doi.org/10.5281/zenodo.6817129.

---

# Supporting Information: Origin of subgap states in normal-insulator-superconductor van der Waals heterostructures

Paritosh Karnatak[1], Zarina Mingazheva[1], Kenji Watanabe[2], Takashi Taniguchi[2], Helmuth Berger[3], László Forró[3,4], Christian Schönenberger[1,5]
[1] *Department of Physics, University of Basel, Basel, Switzerland*
[2] *Research Center for Functional Materials, National Institute for Material Science, 1-1 Namiki, Tsukuba 305-0044, Japan*
[3] *Institute of Condensed Matter Physics, Ecole Polytechnique Fédérale de Lausanne, Lausanne, Switzerland*
[4] *Stavropoulos Center for Complex Quantum Matter, Department of Physics, University of Notre Dame, Notre Dame, IN, USA and*
[5] *Swiss Nanoscience Institute, University of Basel, Basel, Switzerland*

## FABRICATION AND MEASUREMENT

The normal-insulator-superconductor (NIS) type planar tunnel junctions are fabricated by dry stacking of $MoS_2$ ($3-5$ layers) or hBN (3 layers) on $NbSe_2$ crystals ($\sim 3$ nm $-20$ nm). Polycarbonate (PC) backed by PDMS and glass is used for the standard pickup method. The stacking is performed in a glovebox in $N_2$ atmosphere with $H_2O$ and $O_2$ levels below 1 ppm. PC is dissolved in dichloromethane. While $MoS_2$ or hBN act as the tunnel barrier as well as prevent $NbSe_2$ from oxidation during device processing the exposure to the ambient is minimised during device processing. Moreover we encapsulate thin $NbSe_2$ ($\sim 3$ nm) from both sides to prevent degradation. Nearly all devices also contain a graphite crystal transferred over $NbSe_2$, which acts as an ohmic contact to $NbSe_2$. Normal contact regions are defined by ebeam lithography and Ti/Au (5 nm/50 nm) is deposited by ebeam evaporation to create the tunnel junctions. The area of the tunnel junctions is typically $\sim 1$ $\mu$m$^2$ to 3 $\mu$m$^2$, unless stated otherwise. All details of device parameters can be found in the device table at the end of the Supporting Information (SI).

The measurements were performed in a $^3$He fridge at a base temperature of $\sim 250$ mK. The electrical lines are filtered by using pi-filters at the breakout box and tapeworm filters at the cold finger. A small ac voltage ($V_{ac} < k_B T$, where $k_B$ is the Boltzmann constant and $T$ is the cryostat temperature) is added to the biasing dc voltage $V$ using a transformer and the lockin amplifier records the output current via an external current-voltage amplifier. Due to a cryostat installation error the actual magnetic fields may be smaller by up to 10 % than reported here.

## SUPPLEMENTARY DATA: SECOND DERIVATIVE

Figure S1 demonstrates the upper branch of the doublet presented in Figure 2(c) of the main text. For this, the second derivative of the differential conductance with respect to the bias voltage was taken. In the main text we indicate bias voltage as $V$, here, bias voltage is indicated both as $V_b$ and $V$.

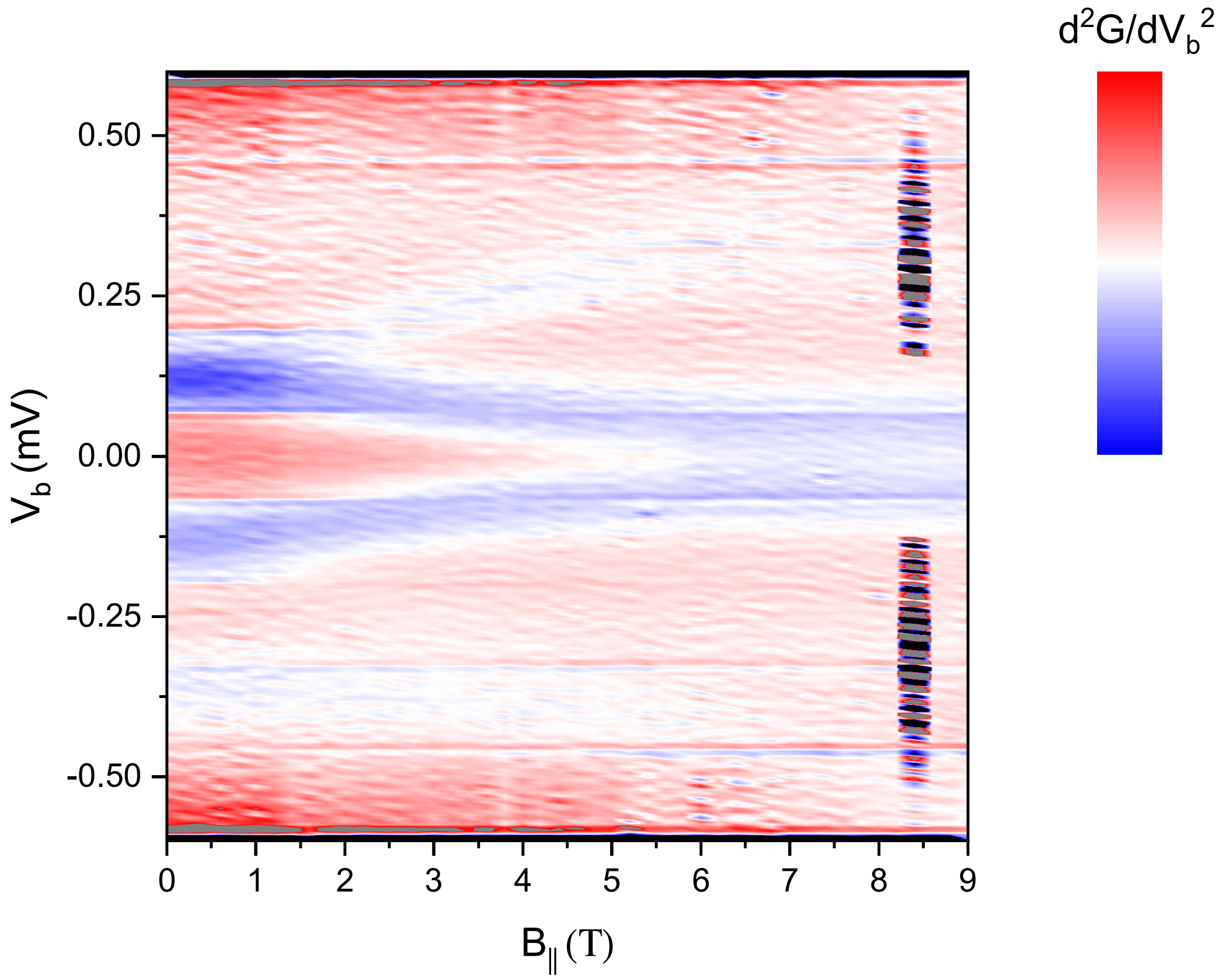

FIG. S1: **2nd derivative of the differential conductance with respect to the bias voltage.** Here G is normalized conductance. Besides branches of doublets which are possible to see in the main text, the second derivative shows clearer the upper branch of the doublet at the positive bias voltage.

## SUPPLEMENTARY DATA: ADDITIONAL SUBGAP SPECTRA

While measuring in the wide range of magnetic field (Figure.S2 and Figure.S3), we take care of conductance jumps due to vortices by shifting dI/dV. The shift was done such that at all magnetic field values, the conductance at $V_b = 0$ has the same value.

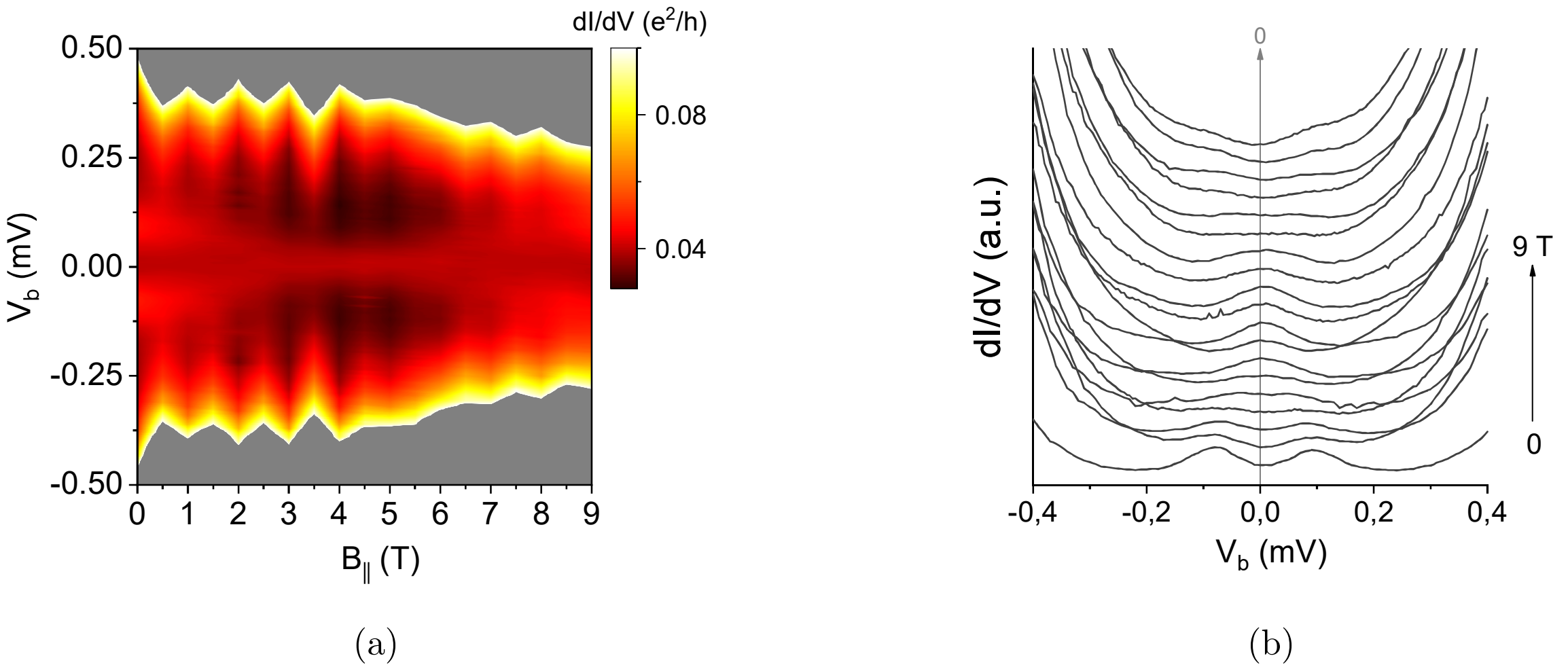

FIG. S2: **Magnetic field evolution of the subgap excitations. This is another example of the transition of the ground state from singlet to doublet under in-plane magnetic field measured for the device D21, tunnel junction no.10.** **a** Colour map of dI/dV as a function of in-plane magnetic field $B_{\|}$ and bias voltage $V_b$. **b** Shifted differential conductance curves for the same tunnel junction as in (**a**). The magnetic field step size of the curves is 0.5 T.

## SUPPLEMENTARY DATA: ADDITIONAL SUBGAP SPECTRA

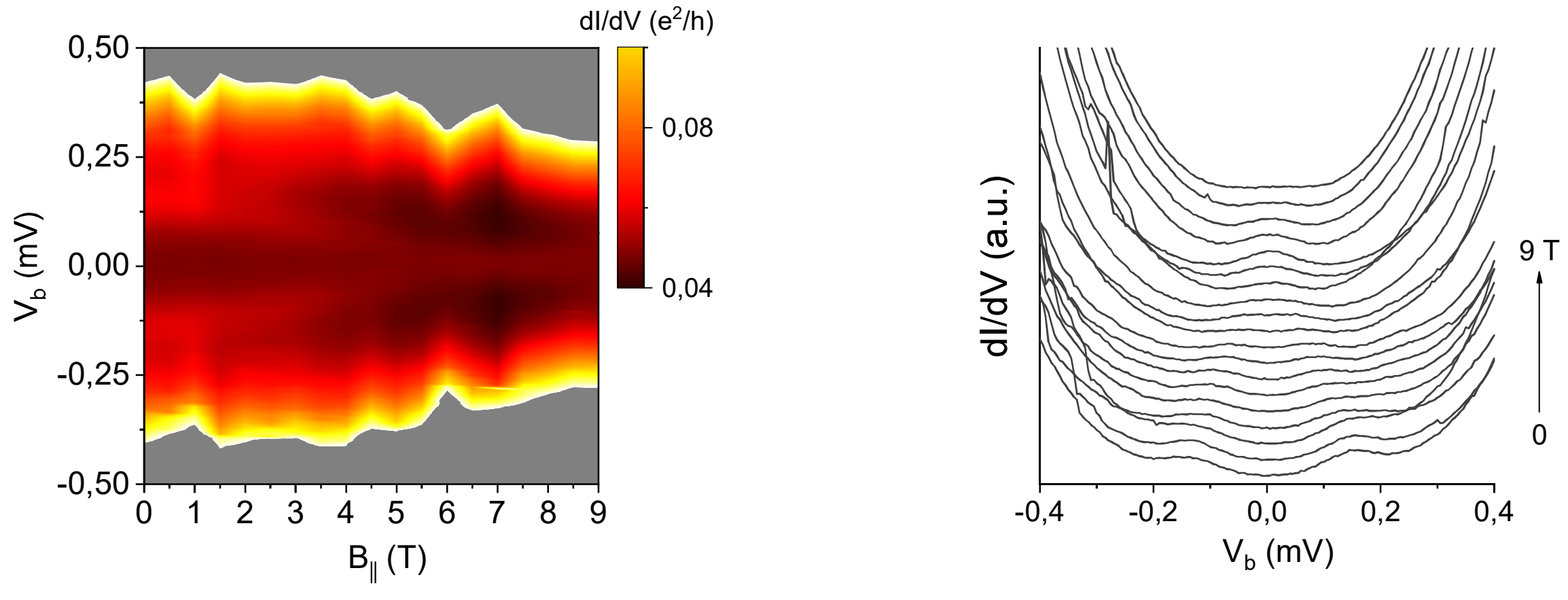

(a) (b)

FIG. S3: **Magnetic field evolution of the subgap excitations, device D21, tunnel junction no.13. a** Colour map of dI/dV as a function of in-plane magnetic field $B_{||}$ and bias voltage $V_b$. **b** Shifted differential conductance curves for the same tunnel junction as in (**a**). The magnetic field step size of the curves is 0.5 T. At $B_{||} = 9$ T, the full width at half maximum (FWHM) is around 0.5 meV. Such a wide peak hinder us to distinguish if the lines cross as in Figure S2a or if the dI/dV peak sticks to $V_b$=0 above 6 T.

## DEVICE DETAILS

The table in this section lists the details of all measured tunnel junctions, particularly geometrical parameters and the suppression factor of the superconducting gap $G_N/G_0$. The main focus is on the type of ground state and its evolution under an in-plane magnetic field (g-factor). In the column "Figure", we indicated in what figures the junction was presented.

A count of the ground states shown in Figure S4 summarized our measurements. More than half of tunnel junctions show subgap excitations, however many of them do not show clear behaviour under an in-plane magnetic field or suppress under a small magnetic field. From the behaviour of the rest of the tunnel junctions, we believe six tunnel junctions have singlet ground state, one - doublet, and four tunnel junctions show accidental degeneracy of a doublet and a hybridized singlet states.

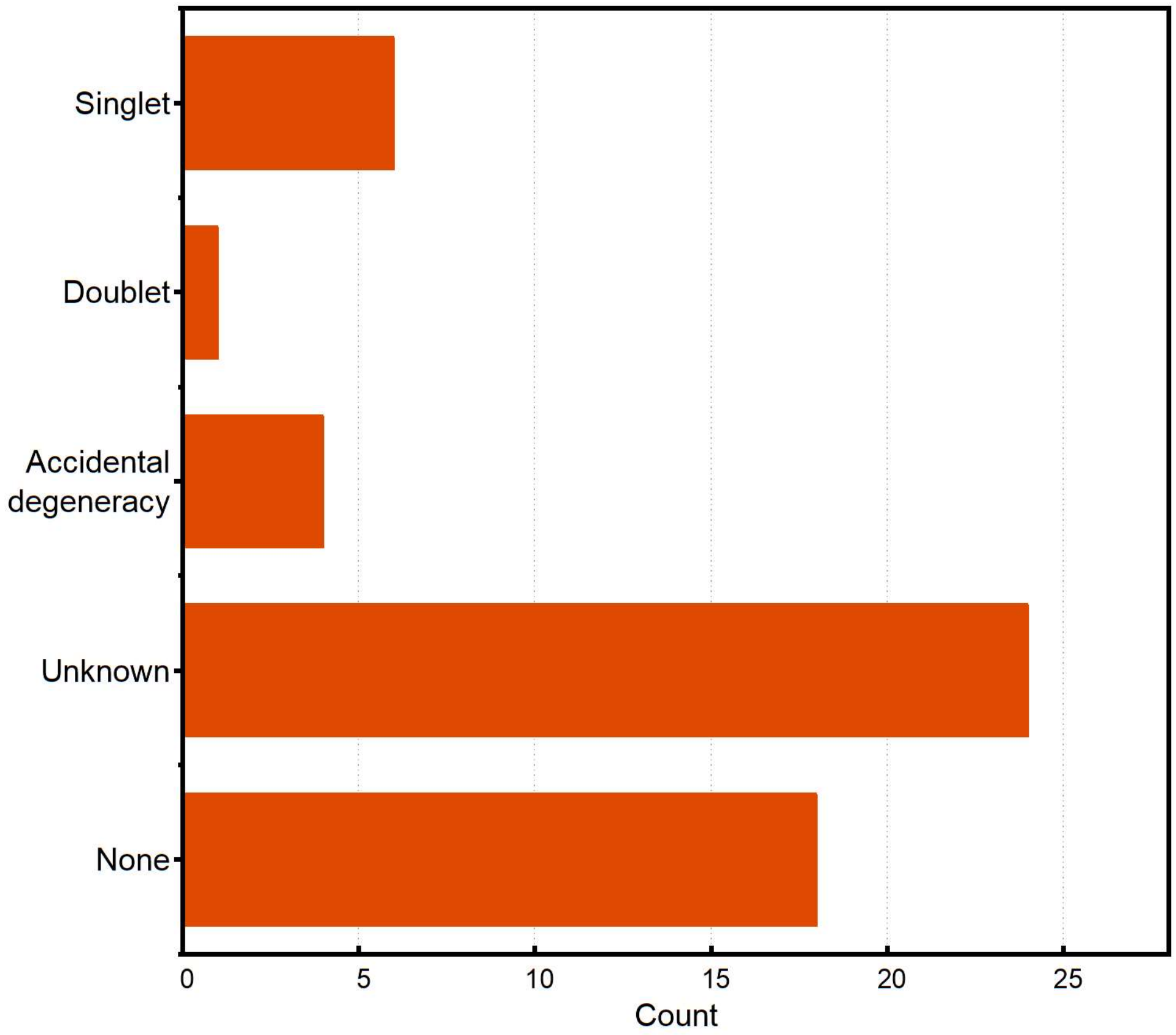

FIG. S4: **Count of ground states of observed in-gap Andreev bound states**

| Details of all measured tunnel junctions (53 junctions) | | | | | | | | | |
|---|---|---|---|---|---|---|---|---|---|
| **Device** | **Junction** | **$NbSe_2$ thickness, nm** | **Tunnel barrier, layers (L)** | **Area of Tj um$^2$** | **$G_N/G_0$** | **Ground state** | **g-factor** | **Comments** | **Figure** |
| D10 | 10 | 3 | 3L $MoS_2$ | 3.2 | 11 | Singlet | 0.70 | | 2a, 2c, S1 |
| | 15 | 3 | 4L $MoS_2$ | 1.1 | 49 | Unknown | 0.00 | Subgap states at the gap edge | 2a |
| | 16 | 7 | 2L $MoS_2$ | 2.6 | 4 | Singlet | 0.70 | Approximate g factor | 2a |
| | 17 | 7 | 4L $MoS_2$ | 2.7 | 400 | None | | | 2a |
| D14 | 1 | 12 | 4L $MoS_2$ | 4 | 63 | Unknown | | on $NbSe_2$ edge | |
| | 5 | 12 | 4L $MoS_2$ | 2.4 | 328 | Singlet | 0.80 | Anticrossing like feature | 3c |
| | 7 | 12 | 4L $MoS_2$ | 3.5 | 100 | Doublet | 0.64 | Higher excitation has a vanishing g-factor | 3a |
| | 14 | 5 | 3L $MoS_2$ | 1.3 | 1 | Unknown | | On $NbSe_2$ edge | |
| | 15 | 10 | 3L $MoS_2$ | 2.3 | 24 | Unknown | | | |
| | 16 | 10 | 3L $MoS_2$ | 2.4 | 4 | Unknown | | | |
| | 17 | 10 | 3L $MoS_2$ | 4.1 | 3 | Accidental degeneracy | 1.70 | | |
| | 18 | 10 | 3L $MoS_2$ | 4.2 | 4 | Unknown | 0.00 | | |
| | 19 | 10 | 3L $MoS_2$ | 1.3 | 44 | Unknown | | | |
| D19 | 3 | 7 | 3L hBN | 10 | 826 | None | | | 2f black |
| | 5 | 7 | 3L hBN | 9.8 | 807 | None | | | 2f red |
| | 7 | 7 | 3L hBN | 10 | 1563 | None | | | 2f blue |
| | 10 | 7 | 3L hBN | 10 | 571 | None | | | 2f green |
| | 11 | 7 | 3L hBN | 10 | 760 | None | | | 2f violet |
| | 16 | 7 | 3L hBN | 12.2 | 500 | None | | | 2f yellow |
| D20 | 1 | 20 | 3L $MoS_2$ | 2.7 | 300 | Unknown | | $NbSe_2$ edge isolated by $MoS_2$ | |
| | 2 | 20 | 3L $MoS_2$ | 2.7 | 150 | Unknown | | $NbSe_2$ edge isolated by $MoS_2$ | 2e black |
| | 3 | 20 | 3L $MoS_2$ | 2.3 | 167 | Unknown | | $NbSe_2$ edge isolated by $MoS_2$ | |
| | 4 | 20 | 3L $MoS_2$ | 2.5 | 80 | Accidental degeneracy | 0.55 | $NbSe_2$ edge isolated by $MoS_2$ | 2e red |

| **Device** | **Junction** | **$NbSe_2$ thickness, nm** | **Tunnel barrier, layers (L)** | **Area of Tj $um^2$** | **$G_N/G_0$** | **Ground state** | **g-factor** | **Comments** | **Figure** |
|---|---|---|---|---|---|---|---|---|---|
| D20 | 5 | 20 | 3L $MoS_2$ | 2.3 | 10 | None | | | |
| | 6 | 20 | 3L $MoS_2$ | 3.1 | 27 | Singlet | 0.45 | Doublet upper branch not visible | |
| | 7 | 20 | 3L $MoS_2$ | 3 | 242 | None | | | |
| | 8 | 20 | 3L $MoS_2$ | 2.9 | 80 | None | | | |
| | 9 | 10 | 3L $MoS_2$ | 1.8 | 131 | None | | | |
| | 10 | 10 | 3L $MoS_2$ | 1.8 | 410 | None | | | |
| | 11 | 10 | 3L $MoS_2$ | 3.2 | 141 | None | | | |
| | 12 | 10 | 3L $MoS_2$ | 1.8 | 8 | None | | V-shaped gap | |
| | 13 | 10 | 3L $MoS_2$ | 2.5 | 2 | Unknown | | V-shaped gap | |
| | 14 | 10 | 3L $MoS_2$ | 2.9 | 104 | None | | | |
| | 16 | 10 | 3L $MoS_2$ | 3.5 | 344 | None | | | |
| D21 | 10 | 8 | 4L $MoS_2$ | 1.52 | 408 | Singlet | 0.75 | | 1d, S2 |
| | 13 | 6 | 4L $MoS_2$ | 1.1 | 425 | Singlet | 0.67 | | S3 |
| | 12 | 8 | 4L $MoS_2$ | 0.71 | 150 | Accidental degeneracy | 0.64 | On $NbSe_2$ edge | 3b |
| | 6 | 8 | 4L $MoS_2$ | 1.96 | 554 | Accidental degeneracy | | | |
| | 7 | 8 | 4L $MoS_2$ | 1.6 | 587 | Unknown | | | |
| | 9 | 8 | 4L $MoS_2$ | 1.96 | 303 | Unknown | | | |
| | 16 | 6 | 4L $MoS_2$ | 1.25 | 630 | None | | | |
| | 11 | 8 | 4L $MoS_2$ | 1.56 | 650 | Unknown | | | |
| | 20 | 6 | 4L $MoS_2$ | 1.4 | 920 | Unknown | | | |
| | 14 | 6 | 4L $MoS_2$ | 1.49 | 271 | None | | | |
| MN1 | 7 | 11 | 3L $MoS_2$ | 1.56 | 826 | Unknown | | Partly located on the hole of $MoS_2$ | 1b,1c |
| | 17 | 11 | 3L $MoS_2$ | 0.68 | 6 | Unknown | | On $NbSe_2$ edge | 2d blue |
| MN4 | 16 | 8 | 5L $MoS_2$ | 3 | 1013 | Unknown | | | |
| | 15 | 8 | 5L $MoS_2$ | 8 | 54 | Unknown | | on $NbSe_2$ crystal step | |
| | 14 | 8 | 4L $MoS_2$ | 9.1 | 334 | Unknown | | on $NbSe_2$ crystal step | |

| **Device** | **Junction** | **$NbSe_2$ thickness, nm** | **Tunnel barrier, layers (L)** | **Area of Tj um$^2$** | **$G_N/G_0$** | **Ground state** | **g-factor** | **Comments** | **Figure** |
|---|---|---|---|---|---|---|---|---|---|
| MN4 | 11 | 10 | 4L $MoS_2$ | 4.6 | 359 | Unknown | | | |
| | 3 | 14 | 5L $MoS_2$ | 3.5 | 327 | Unknown | | on $NbSe_2$ edge | |
| | 9 | 10 | 5L $MoS_2$ | 6.2 | 52 | Unknown | | on $NbSe_2$ edge | 2d black and red |
| | 17 | 8 | 5L $MoS_2$ | 3.64 | 145 | Unknown | | Subgap states at the gap edge | |